\documentclass[preprint,12pt]{elsarticle}

\usepackage{amsmath,amssymb}
\usepackage{graphicx}
\usepackage{lineno}
\usepackage{bm}
\usepackage{siunitx}
\usepackage{color}

\journal{Current Applied Physics}

\begin{document}

\begin{frontmatter}

\title{Asymmetric-gate Mach--Zehnder interferometry in graphene: Multi-path conductance oscillations and visibility characteristics}

\author[inst1]{Taegeun Song}
\author[inst2,inst3]{Nojoon Myoung\corref{cor1}}
\ead{nmyoung@chosun.ac.kr}

\cortext[cor1]{Corresponding author}
\address[inst1]{Department of Data Information and Physics, Kongju University, Kongju 32588, Republic of Korea}
\address[inst2]{Department of Physics Education, Chosun University, Gwangju 61452, Republic of Korea}
\address[inst3]{Institute of Well-Aging Medicare \& CSU G-Lamp Project Group, Chosun University, Gwangju 61452, Republic of Korea}

\begin{abstract}
Graphene provides an excellent platform for investigating electron quantum interference due to its outstanding coherent properties. In the quantum Hall regime, Mach--Zehnder (MZ) electronic interferometers are realized using p--n junctions in graphene, where electron interference is highly protected against decoherence. In this work, we present a phenomenological framework for graphene-based MZ interferometry with asymmetric p--n junction configurations. We show that the enclosed interferometer area can be tuned by asymmetric gate potentials, and additional MZ pathways emerge in higher-filling-factor scenarios, e.g. $\left(\nu_{n},\nu_{p}\right)=\left(-3,+3\right)$. The resulting complicated beat oscillations in asymmetric-gate MZ interference are efficiently analyzed using a machine-learning--based Fourier transform, which yields improved peak-to-background ratios compared to conventional signal-processing techniques. Furthermore, we examine the impact of the asymmetric gate on the interference visibility, finding that interference visibility is enhanced under symmetric gate conditions.
\end{abstract}


\end{frontmatter}

\section{Introduction}
Electron quantum optics has emerged from a direct analogy between phase-coherent electron transport in mesoscopic circuits and photon propagation in optical systems. Phase-coherent currents in electron quantum interferometers exemplify electron quantum optic phenomena. Graphene's high mobility, zero bandgap, and tunable carrier type make it an ideal platform for electron quantum optics; in recent years, graphene-based interferometers have significantly advanced our understanding of phase-coherent electronic currents\cite{chakraborti2025electron,jo2021quantum,jo2022scaling,wei2017mach,rickhaus2013ballistic,deprez2021tunable,dauber2017aharonov}.

In particular, in the quantum Hall regime, a graphene Mach--Zehnder (MZ) interferometer is realized at an electrostatically defined p--n junction under a strong perpendicular magnetic field, where one-dimensional chiral edge channels serve as coherent pathways, thereby protecting interference against decoherence\cite{wei2017mach,matsuo2015edge,morikawa2015edge,zimmermann2017tunable}. Compared to GaAs edge channel interferometers, graphene's zero bandgap enables seamless coexistence and manipulation of hole- and electron-like modes at the p--n interface\cite{dubey2016tuning,klimov2015edge,jo2021quantum}. Moreover, control of valley isospin offers a route to engineer interchannel scattering---an essential ingredient for reconfigurable edge-channel interferometry\cite{tworzydlo2007valley,trifunovic2019valley,myoung2020manipulation,handschin2017giant,jo2021quantum}.

From an application point of view, based on insights into quantum interference in graphene-based MZ interferometers, sensing applications have been proposed to detect physical quantities such as spin waves and local deformation\cite{myoung2024detecting,assouline2021excitonic,wei2018electrical}. In a prototypical graphene MZ interferometry operating in the quantum Hall regime, an electron beam is split into two coherent paths at the first beam splitter and recombined at the second. The interferometry's sensing quality is governed by the splitting ratio: equal partitioning yields maximal visibility between constructive and destructive interference. In principle, this splitting ratio depends on valley scattering processes at the p--n junction, where electron- and hole-like quantum Hall edge channels merge\cite{jo2021quantum,tworzydlo2007valley}. Another factor affecting visibility is the spatial separation between these quantum Hall channels along the junction, which is determined by the smoothness of the gate-defined junction potential\cite{myoung2020manipulation}. Because the positions of quantum Hall channels at the p--n interface can be tuned via the gate potential\cite{handschin2017giant}, understanding how the junction potential profile influences quantum interference is crucial for optimizing graphene MZ interferometry--based quantum sensors.

In fact, when the filling factors in the p and n regions are set to $\left(\nu_{p},\nu_{n}\right)=\left(-1,1\right)$, MZ interference vanishes in the graphene p--n junction. However, as soon as one or both filling factors shift to $\pm3$, clear conductance oscillations appear. This behavior stems from the spatial merging of the $\nu_{n,p}=\pm1$ quantum Hall channels at the junction, which prevents coherent path separation\cite{myoung2020manipulation}. Experimental studies have demonstrated markedly different visibilities for configurations $\left(\nu_{p},\nu_{n}\right)=\left(-1,3\right)$ versus $\left(\nu_{p},\nu_{n}\right)=\left(-3,3\right)$\cite{wei2017mach,jo2022scaling}. However, despite these observations, a detailed understanding of how gate-defined potential profiles modulate interference visibility remains lacking.

In this work, we develop a phenomenological framework to elucidate how an asymmetric gate potential affects quantum interference in graphene MZ interferometry. We combine numerical simulations with machine-learning--based signal analysis of conductance oscillations, achieving an improved peak-to-background ratio over conventional methods, and show that complete MZ pathways---comprising two spatially isolated quantum Hall channels at the p--n interface---are essential for interference to occur. When higher Landau levels contribute one or two additional pairs of interface channels, multi-path interference emerges. We further demonstrate that the effective enclosed area of each MZ loop is tunable via gate-voltage asymmetry, and that beam-splitting efficiency likewise varies with the junction's potential profile. Finally, we confirm that maximal visibility is achieved when the p--n junction is symmetric, highlighting a clear design rule for optimizing graphene MZ-based quantum sensors.

\section{Theoretical Framework}

\begin{figure}[htpb!]
    \centering
    \includegraphics[width=0.5\linewidth]{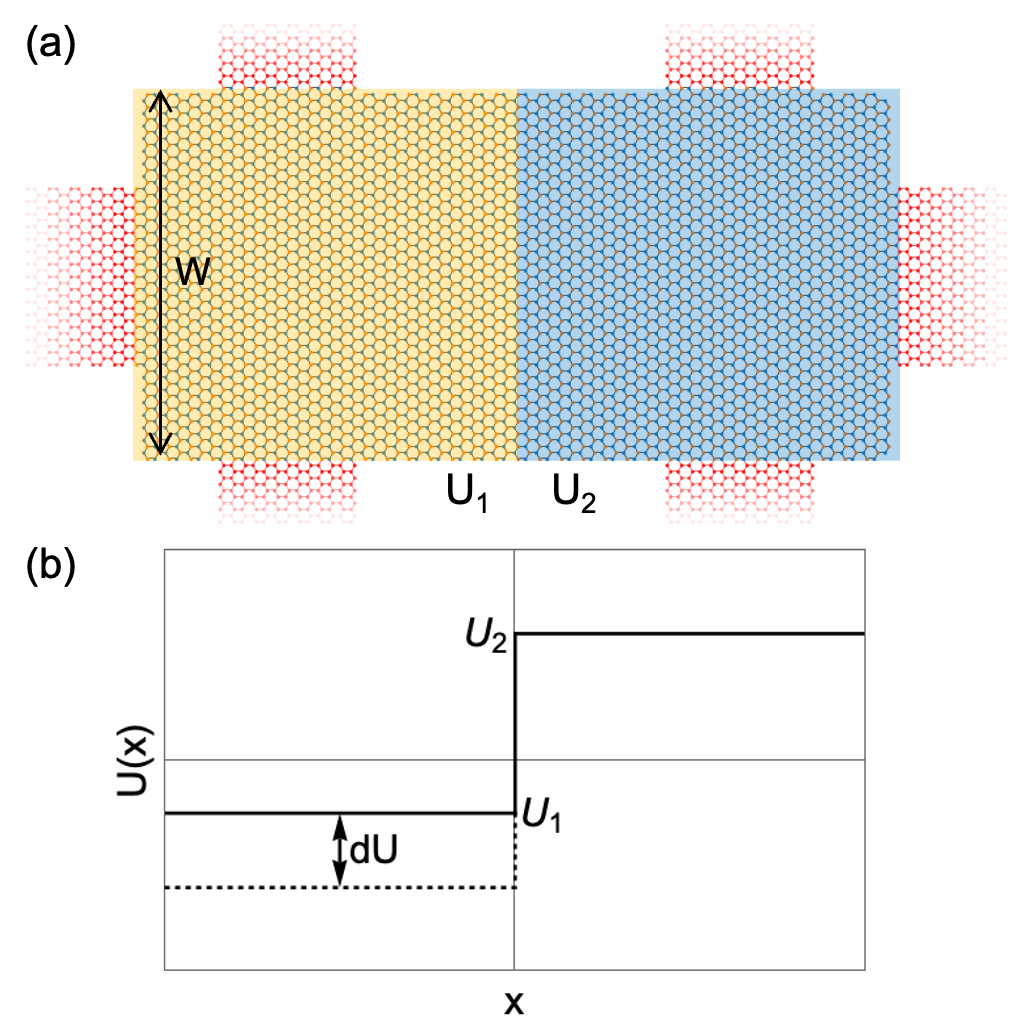}
    \caption{(a) Schematic of the graphene Hall bar with an electrostatically defined p--n junction; edge-channel conductance is measured via two opposite leads attached to the same region. (b) Potential profile for the p--n junction. N and p regions are subjected to electric potential $U_{1}$ and $U_{2}$, respectively. Solid line represents an asymmetric junction potential $U_{1}=-E_{0}/2$ and $U_{2}=E_{0}\left(1+\sqrt{2}\right)/2$ for filling-factor configuration $\left(\nu_{p},\nu_{n}\right)=\left(-1,3\right)$, while dashed line represents represents a symmetric junction potential $U_{1}=-U_{2}=-E_{0}\left(1+\sqrt{2}\right)/2$ for filling-factor configuration $\left(\nu_{p},\nu_{n}\right)=\left(-3,3\right)$.}
    \label{fig:model}
\end{figure}

In the quantum Hall regime, a graphene MZ interferometer is implemented via an electrostatically defined p--n junction on a graphene Hall bar, as shown in Fig. \ref{fig:model}(a). Its low-energy behavior is described by the Dirac Hamiltonian.
\begin{align}
    H=\hbar v_{F}\vec{\sigma}\cdot\left(\vec{k}+e\vec{A}\right)+U\left(x\right),\label{eq:DiracHam}
\end{align}
where $\vec{\sigma}=\left(\sigma_{x},\sigma_{y}\right)$ are the Pauli matrices and $v_{F}\simeq 10^{6}~\mathrm{ms}^{-1}$ is the Fermi velocity of the Dirac fermions. Here, $\vec{A}$ is the magnetic vector potential corresponding to a uniform magnetic field $\vec{B}=B\hat{z}=\vec{\nabla}\times\vec{A}$. By assuming the abrupt-junction limit, the electrostatic potential for the p--n junction is defined as
\begin{align}
    U\left(x\right)=\left\{\begin{array}{ll}U_{1},&x<0,\\U_{2},&x>0.\end{array}\right.,\label{eq:pot}
\end{align}
depicted in Fig. \ref{fig:model}(b). Previous work has shown that a smooth junction reduces interference visibility in the graphene MZ interferometer\cite{myoung2017conductance}.

In the abrupt junction limit, the Dirac equation [Eq. (\ref{eq:DiracHam})] admits analytical solutions in terms of parabolic cylinder function:
\begin{align}
    \psi_{1,2}=A\left(\begin{array}{c}D_{\nu_{1,2}}\left(s\zeta\right)\\
    -is\sqrt{\frac{\nu}{2}}D_{\nu_{1,2}-1}\left(s\zeta\right)\end{array}\right),
\end{align}
where $A$ is the normalization constant, $s=\mathrm{sgn}\left(x\right)$, $\zeta\equiv 2k_{y}-x$, and $\nu_{1,2}=\left(E-sU_{1,2}\right)^{2}$. Applying the continuity of the wavefunction, we have the secular equation
\begin{align}
    \mbox{det}\left[\begin{array}{cc}D_{\nu_{1}}\left(-2k_{y}\right)&-D_{\nu_{2}}\left(2k_{y}\right)\\
    i\sqrt{\frac{\nu_{1}}{2}}D_{\nu_{1}-1}\left(-2k_{y}\right)&i\sqrt{\frac{\nu_{2}}{2}}D_{\nu_{2}-1}\left(2k_{y}\right)\end{array}\right]=0.
\end{align}
Numerically solving this equation for each $k_{y}$, we obtain the eigenenergy dispersion $E\left(k_{y}\right)$.

Transforming the Dirac Hamiltonian in Eq. (\ref{eq:DiracHam}), we obtain a Schr\"{o}dinger-like equation for each spinor componenet $\psi_{A,B}$:
\begin{align}
    \left[\frac{d^{2}}{dx^{2}}+U_{eff}+\left(E-U\right)^{2}\right]\psi_{A,B}=0,
\end{align}
where the effective potential
\begin{align}
    U_{eff}\equiv -\frac{\varsigma}{2}+\left(k_{y}-\frac{x}{2}\right)^{2},
\end{align}
exhibits a local minimnum at $x=2k_{y}$. This effective potential picture provides a quantitative understanding of where the interface states form along the junction.

Defining the Fermi level $E_{F}$ of the system, the number of interface channels is given by the number of intersections between the energy dispersion and $E_{F}$, counted with two-fold valley degeneracy. Equivalently, this channel count equals the Chern-number difference across the p--n junction. Because $U_{eff}$ is quadratic near its minimum, overlap of wavefunctions between adjacent interface channels must undergo an exponential decay as the distance increases. The key question then is: How large is the decay length and how does this decay influence the visibility of the graphene MZ interferometer?

We investigated graphene MZ interference by analyzing oscillations in the quantum hall conductance, which arise from coupling between counter-propagating edge channels. (see Fig. \ref{fig:model}(a)). Even without localized states in the bulk, the interface channels along the p--n interface enable mode-mixing that underpins the interface signal.

Our conductance calculations employ a nearest-neighbor tight-binding Hamiltonian.
\begin{align}
    H_{tb}=\sum_{\left<i,j\right>}t_{ij}\left(c^{\dagger}_{i}c_{j}+c_{j}^{\dagger}c_{i}\right)+\sum_{i}U\left(\vec{r}_{i}\right)c^{\dagger}_{i}c_{i},
\end{align}
where $c_{i}$($c^{\dagger}_{i}$) is an annihilation(creation) operator of an elecrron at site $i$ with on-site potential $U\left(\vec{r}_{i}\right)$ defined in Eq. (\ref{eq:pot}), and its position $\vec{r}_{i}$, and the nearest-neighbor hopping energy reads
\begin{align}
    t_{ij}=t_{0}~\mathrm{exp}\left(i\frac{e}{\hbar}\int_{\vec{r}_{j}}^{\vec{r}_{i}}\vec{A}\cdot d\vec{r}\right),
\end{align}
where $t_{0}$ is the zero-field hopping amplitude. Using \textsc{kwant} code for tight-binding scattering-matrix calculations\cite{groth2014kwant}, we compute the multi-terminal quantum Hall conductance via Landauer--B\"{u}ttiker formula:
\begin{align}
    G_{H}=\frac{2e^{2}}{h}\sum_{m\in{\alpha}}\sum_{n\in{\beta}}\left|S_{nm}\right|^{2},
\end{align}
where $S_{nm}$ is the scattering-matrix element from mode $m$ in lead $\beta$ to mode $n$ in lead $\alpha$.

To analyze the complicated beat patterns in our conductance oscillations with high resolution, we employ a machine learning–based Fourier transform (ML-FT) approach. This technique provides significantly higher frequency resolution than a conventional Fast Fourier Transform, enabling the extraction of precise frequency components even from relatively short signal segments. In other words, closely spaced frequencies that produce complex beating can be distinguished as separate, clear peaks in the spectrum. A distinct advantage of this ML-FT method is its ability to retrieve accurate frequency information even when the available data length is limited – a regime where standard FFT analysis often falters. This makes the ML-based approach especially well-suited for our study, allowing us to disentangle multiple interference frequencies with improved clarity and confidence. 

In practice, the ML-FT is implemented using a single-hidden-layer artificial neural network with a sinusoidal activation function in the hidden layer. We use 1000 hidden nodes in this layer, providing a rich basis set for representing the oscillatory signal. The output layer is linear, summing contributions from all hidden nodes. With this architecture, each hidden neuron effectively learns one sinusoidal component of the input signal, so the network as a whole produces a Fourier-like decomposition of the data (details of this sinusoidal network approach are provided elsewhere \cite{myoung2024detecting}). We trained the network on the conductance oscillation data using the Adam optimizer with a mean-squared-error loss, while employing a cosine-annealing learning rate schedule for efficient convergence. Additionally, we applied weight clipping to the input-to-hidden layer weights (which determine the sine frequencies) to confine the search to the relevant frequency range. After training, the learned weight parameters directly yield the spectrum of the signal: the frequency of each hidden neuron’s sine corresponds to a detected frequency component, and its associated output weight gives the amplitude. Notably, the resulting ML-FT spectrum consists of extremely sharp, delta-like peaks at the learned frequencies, resulting in a high peak-to-background ratio. This sharpness greatly facilitates interpretation of the results, as even weak or closely adjacent frequency components appear as well-resolved peaks. Indeed, the “ML-FT amplitude” spectra obtained by this method (see Figure 2 and Figure 3) exhibit distinct narrow peaks corresponding to the fundamental and beat frequencies, consistent with the enhanced resolution and accuracy afforded by the ML-based approach. 

\section{Results and Discussion}

\begin{figure*}[hptb!]
    \centering
    \includegraphics[width=\linewidth]{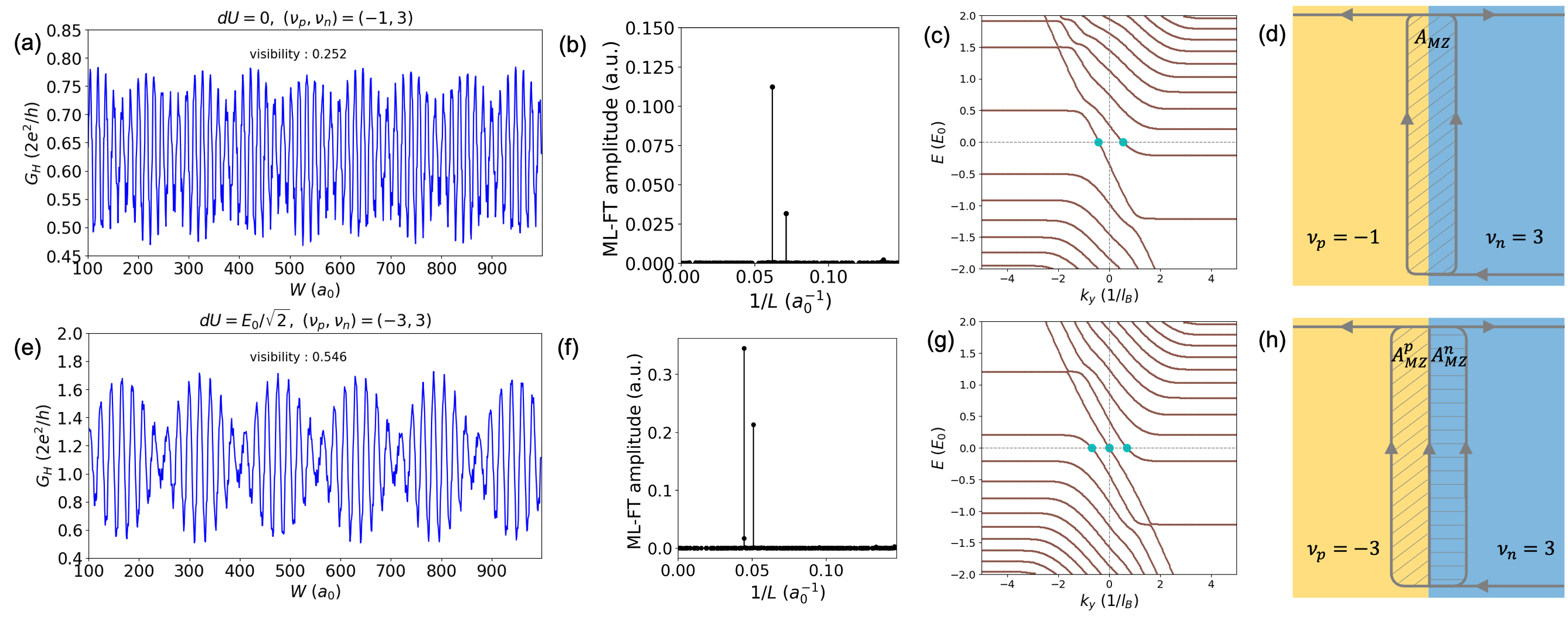}
    \caption{Conductance oscillations, machine‐learning–based Fourier analysis results, and corresponding interface‐channel dispersions and interferometer loops for two gate‐asymmetry settings. (a) Quantum Hall conductance $G_{H}$ versus junction length $L$ for the asymmetric p--n junction $\left(dU=0\right)$ at a filling-factor configuration $\left(\nu_{p},\nu_{n}\right)=\left(-1,3\right)$; the visibility of the oscillations is indicated. (b) Machine-learning Fourier transform of the conductance oscillation in (a), showing the valley-split MZ frequencies. (c) Energy spectrum of the interface channels for $\left(\nu_{p},\nu_{n}\right)=\left(-1,3\right)$; solid dots mark the two modes $k_{0}$ and $k_{+1}$ crossing the Fermi energy $E=0$. (d) Schematic of the corresponding Mach-Zehnder interfereometer loop: the shaded area $A_{\rm MZ}$ enclosed by the two counter-propagating edge channels in the p– and n–regions encloses magnetic flux $\Phi=B\cdot A_{\rm MZ}$. (e) Conductance oscillations for the symmetric junction potential $dU=E_{0}/\sqrt{2}$ at $(\nu_{p},\nu_{n})=(-3,3)$; the visibility is noted. (f) Machine‐learning Fourier transform of the oscillations in (e), with the valley-split MZ frequencies. (g) Dispersion of the interface channels for $(\nu_{p},\nu_{n})=(-3,3)$; three Fermi‐level crossings indicate three interface modes $k_{-1}$, $k_{0}$ and $k_{+1}$, respectively. (h) Schematic of the nested Mach–Zehnder loops $A^{p}_{\rm MZ}$ and $A^{n}_{\rm MZ}$ formed by the two pairs of interface channels in the higher–filling–factor case.}
    \label{fig:MZI}
\end{figure*}

First, we compare conductance oscillations of the graphene MZ interferometer in asymmetric and symmetric gate-potential configurations [Fig. \ref{fig:MZI}(a) and (e)]. Both cases exhibit clear beat patterns as the channel length $L$ varies. As shown previously\cite{myoung2017conductance}, these beat oscillations arise from the valley-split quantum Hall interface channels that enclose a slightly different area of the MZ-interferometer.

Our ML-FT reveals two distinct peaks at $f_{1}=0.255~\mathrm{nm^{-1}}$ and $f_{1}'=0.296~\mathrm{nm^{-1}}$. Since the interface channels reside at the local minima of $U_{eff}\left(x\right)$, the width of the MZ loop is $\Delta x=2\left(k_{+1}-k_{0}\right)l_{B}=1.136~\mathrm{nm}$, where $k_{+1}$ and $k_{0}$ are the Fermi-energy crossing indicated in Fig. \ref{fig:MZI}(c). This value $\Delta x$ predicts a first-harmonic frequency $eB\Delta x/h=0.259~\mathrm{nm^{-1}}$, in excellent agreement with the ML-FT result. The difference in the loop widths between two valley-resolved channels is very small, about $0.179~\mathrm{nm}$. For the symmetric junction case, we also find that the analytically estimated and ML-based results of the first-harmonic MZ frequency match closely.

\begin{figure}[htpb!]
    \centering
    \includegraphics[width=0.5\linewidth]{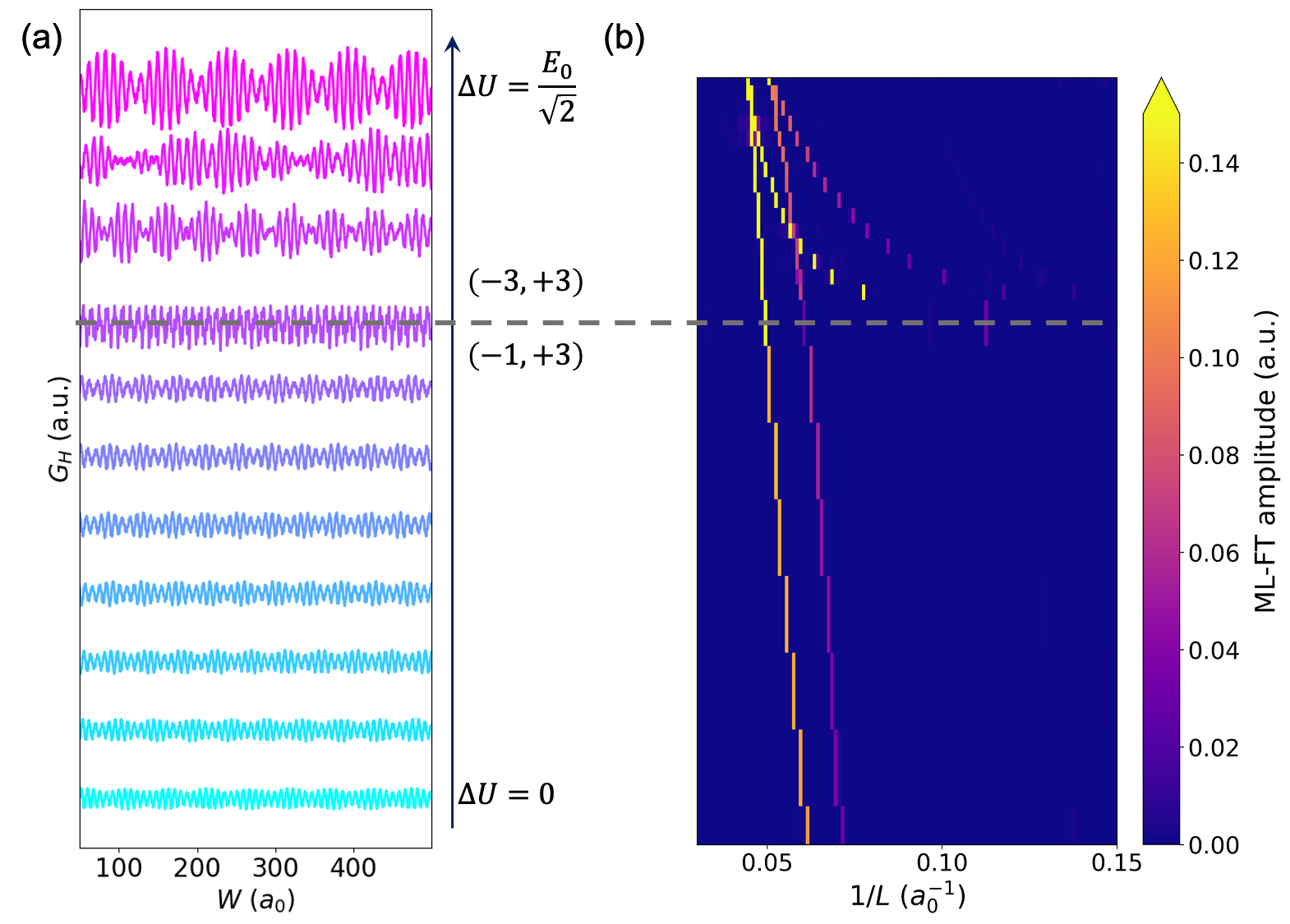}
    \caption{Evolution of graphene Mach-Zehnder interference with varying p--n junction asymmetry. (a) Stacked plots of the quantum Hall conductance $G_{H}$ versus junction length for a series of potential-difference values $dU$. The arrow at right indicates the direction of increasing $dU$. The dashed horizontal line marks the singular filling-factor case that one region of p--n junction is nearly metallic (Fermi energy touches the first Landu level). (b) ML-FT spectra computed from each trace in (a), illustrating the evolution of the dominant beat frequencies and the emergence of additional frequency components as $dU$ grows.}
    \label{fig:MLFourier}
\end{figure}

Next, we examine how the $G_{H}$ oscillation spectra evolve with the p--n junction asymmetry. Figure \ref{fig:MLFourier} presents the conductance trace along with their ML-FT for various $dU$ values. For the $\left(\nu_{p},\nu_{n}\right)=\left(-1,3\right)$ configuration, the primary interface frequency continuously shifts as $dU$ increases, reflecting the narrowing MZ loop width. In contrast, the $\left(-3,3\right)$ configuration exhibits the emergence of additional frequency components corresponding to nested loops. As $dU$ approaches $E_{0}/\sqrt{2}$ (the symmetric junction), these multiple peaks coalesce into a single frequency, since the MZ loops in the p and n regions enclose identical areas. In other words, under symmetric gating, two equivalent MZ loops contribute equally to the beat pattern in the $G_{H}$ spectra. Finally, although a higher-order loop enclosing area $A_{MZ}^{n}+A_{MZ}^{p}$ can generate secondary harmonics of the MZ interference, its Fourier-peak amplitude remains negligible compared to the primary modes (see the residual peak in Fig. \ref{fig:MLFourier}(b)). 

We note that at $dU=0.7E/\sqrt{2}$, the n-region potential $U_{1}=-E_{0}/2 + dU$ becomes nearly $-E_{0}$, causing the first Landau level to approach the Fermi energy of the system. Because the existence of topologically protected interface channels requires both regions to be in a Chern-insulating state, the junction at $dU=0.7E_{0}/\sqrt{2}$ lies in a topological singularity. For this reason, we excluded the $G_{H}$ spectrum at $dU=0.7E_{0}/\sqrt{2}$ from both the ML-FT analysis and the visibility examination. 

\begin{figure}[htpb!]
    \centering
    \includegraphics[width=0.5\linewidth]{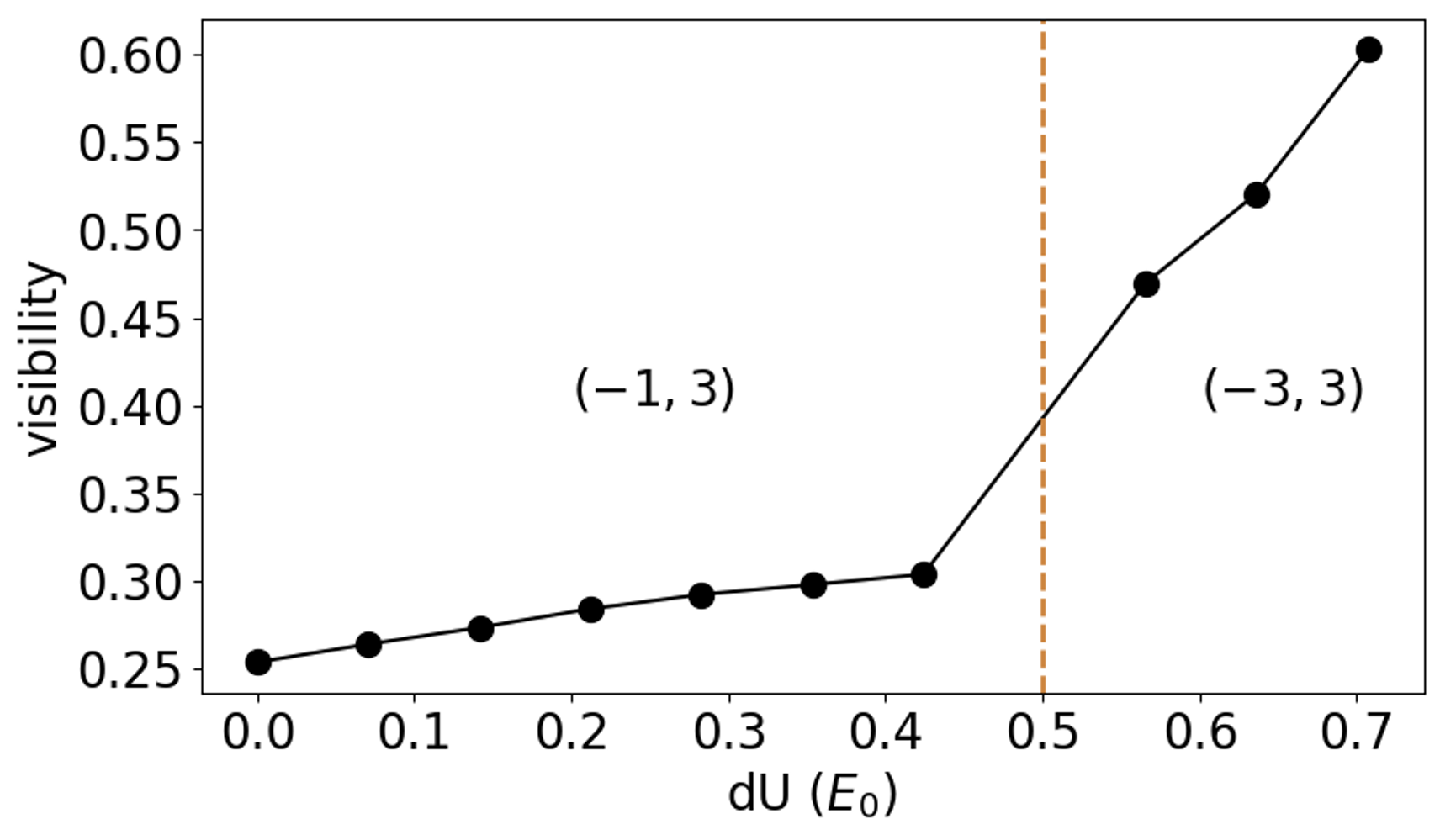}
    \caption{Dependence of MZ interference visibility on p--n junction asymmetry $dU$. Visibility is extracted from conductance oscillations (Fig. \ref{fig:MLFourier}(a)) for filing-factor configurations $\left(\nu_{p},\mu_{n}\right)=\left(-1,3\right)$ at small $dU$ and $\left(\nu_{p},\mu_{n}\right)=\left(-3,3\right)$ once $dU$ exceeds the threshold (vertical dashed line at $dU=E_{0}/2$. Each marker denotes the calculated visibility for a given $dU$. The sharp increase in visibility beyond $dU=E_{0}/2$ reflects the transition from single-loop to nested multi-path MZ interference.}
    \label{fig:visibility}
\end{figure}

Finally, we assess how junction asymmetry affects interference visibility. We define visibility as
\begin{align}
    V=\frac{G_{H,\mathrm{max}}-G_{H,\mathrm{min}}}{G_{H,\mathrm{max}}+G_{H,\mathrm{min}}}.
\end{align}
Comparing the conductance oscillations in Figs. \ref{fig:MZI}(a) and (e), it is evident that the $\left(\nu_{p},\nu_{n}\right)=\left(-3,3\right)$ configuration yields a higher visibility than the $\left(-1,3\right)$ case. As noted above, the visibility depends on the separation between the interface channels that bound the MZ loop. In the $\left(-1,3\right)$ configuration, the enclosed area $A_{MZ}=\Delta x_{MZ}\cdot L$ exceeds that of each individual loop in the symmetric $\left(-3,3\right)$ case, where $A_{MZ}^{n}=A_{MZ}^{p}=\Delta x_{MZ}^{p}\cdot L$. Consequently, symmetric gating improves beam-splitting efficiency and enhances visibility. Figure \ref{fig:visibility} summarizes this trend by plotting $V$ versus $dU$, showing the monotonic increase in visibility as the junction potential approaches the symmetric case. 

\section{Conclusions}

Our study demonstrates that the Mach-Zehnder(MZ) interference in graphene quantum Hall p--n junctions depends on the gate-potential configuration. Notably, one-dimensional, dissipationless quantum Hall interface channels provide a tunable medium for coherent single-electron propagation, with trajectories that enclose MZ loops suitable for quantum interferometry. While prior work on graphene p--n junctions in the quantum Hall regime focused on optimizing a specific filling-factor configuration for maximal visibility\cite{wei2017mach,jo2021quantum}, we reveal the interference mechanisms arising from co-propagating channels at different Landau levels under general asymmetric gating.

To quantify these effects, we applied a machine-learning--based Fourier analysis\cite{myoung2024detecting} to the conductance oscillations from graphene MZ interferometry devices. In addition to the standard valley-split MZ interference loop, we observe the emergence of multiple MZ loops at the higher filling-factor configuration $\left(\nu_{p},\nu_{n}\right)=\left(-3,3\right)$, reflecting additional interface channels at the p--n junction. Quantitative and qualitative estimates of magnetic flux through each MZ loop explain the resulting complicated beat patterns in the conductance spectra. We further show that interference visibility decreases under asymmetric gating compared to the symmetric case. These findings lay the groundwork for diverse inteferometric experiments---and offer diagnostic tools capable of detecting subnanometer-scale shifts in interface-channel separation at the p--n junction, enabling sensitive quantum interferometric sensing. 

\section*{Acknowledgments}
This work was supported by Chosun University(2024)

\section*{Declaration of Competing Interest}
The authors declare that they have no known competing financial interests or personal relationships that could have appeared to influence the work reported in this paper.

\bibliographystyle{elsarticle-num}

\bibliography{AsymVisMZGra}

\begin{thebibliography}{10}
\expandafter\ifx\csname url\endcsname\relax
  \def\url#1{\texttt{#1}}\fi
\expandafter\ifx\csname urlprefix\endcsname\relax\def\urlprefix{URL }\fi
\expandafter\ifx\csname href\endcsname\relax
  \def\href#1#2{#2} \def\path#1{#1}\fi

\bibitem{chakraborti2025electron}
H.~Chakraborti, L.~Pugliese, A.~Assouline, K.~Watanabe, T.~Taniguchi, N.~Kumada, D.~Glattli, M.~Jo, H.-S. Sim, P.~Roulleau, Electron collision in a two-path graphene interferometer, Science 388~(6746) (2025) 492--496.

\bibitem{jo2021quantum}
M.~Jo, P.~Brasseur, A.~Assouline, G.~Fleury, H.-S. Sim, K.~Watanabe, T.~Taniguchi, W.~Dumnernpanich, P.~Roche, D.~Glattli, et~al., Quantum hall valley splitters and a tunable mach-zehnder interferometer in graphene, Physical Review Letters 126~(14) (2021) 146803.

\bibitem{jo2022scaling}
M.~Jo, J.-Y.~M. Lee, A.~Assouline, P.~Brasseur, K.~Watanabe, T.~Taniguchi, P.~Roche, D.~Glattli, N.~Kumada, F.~Parmentier, et~al., Scaling behavior of electron decoherence in a graphene mach-zehnder interferometer, Nature Communications 13~(1) (2022) 5473.

\bibitem{wei2017mach}
D.~S. Wei, T.~van~der Sar, J.~D. Sanchez-Yamagishi, K.~Watanabe, T.~Taniguchi, P.~Jarillo-Herrero, B.~I. Halperin, A.~Yacoby, Mach-zehnder interferometry using spin-and valley-polarized quantum hall edge states in graphene, Science advances 3~(8) (2017) e1700600.

\bibitem{rickhaus2013ballistic}
P.~Rickhaus, R.~Maurand, M.-H. Liu, M.~Weiss, K.~Richter, C.~Sch{\"o}nenberger, Ballistic interferences in suspended graphene, Nature communications 4~(1) (2013) 2342.

\bibitem{deprez2021tunable}
C.~D{\'e}prez, L.~Veyrat, H.~Vignaud, G.~Nayak, K.~Watanabe, T.~Taniguchi, F.~Gay, H.~Sellier, B.~Sac{\'e}p{\'e}, A tunable fabry--p{\'e}rot quantum hall interferometer in graphene, Nature nanotechnology 16~(5) (2021) 555--562.

\bibitem{dauber2017aharonov}
J.~Dauber, M.~Oellers, F.~Venn, A.~Epping, K.~Watanabe, T.~Taniguchi, F.~Hassler, C.~Stampfer, Aharonov-bohm oscillations and magnetic focusing in ballistic graphene rings, Physical Review B 96~(20) (2017) 205407.

\bibitem{matsuo2015edge}
S.~Matsuo, S.~Takeshita, T.~Tanaka, S.~Nakaharai, K.~Tsukagoshi, T.~Moriyama, T.~Ono, K.~Kobayashi, Edge mixing dynamics in graphene p--n junctions in the quantum hall regime, Nature communications 6~(1) (2015) 8066.

\bibitem{morikawa2015edge}
S.~Morikawa, S.~Masubuchi, R.~Moriya, K.~Watanabe, T.~Taniguchi, T.~Machida, Edge-channel interferometer at the graphene quantum hall pn junction, Applied Physics Letters 106~(18) (2015).

\bibitem{zimmermann2017tunable}
K.~Zimmermann, A.~Jordan, F.~Gay, K.~Watanabe, T.~Taniguchi, Z.~Han, V.~Bouchiat, H.~Sellier, B.~Sac{\'e}p{\'e}, Tunable transmission of quantum hall edge channels with full degeneracy lifting in split-gated graphene devices, Nature Communications 8~(1) (2017) 14983.

\bibitem{dubey2016tuning}
S.~Dubey, M.~M. Deshmukh, Tuning equilibration of quantum hall edge states in graphene--role of crossed electric and magnetic fields, Solid State Communications 237 (2016) 59--63.

\bibitem{klimov2015edge}
N.~N. Klimov, S.~T. Le, J.~Yan, P.~Agnihotri, E.~Comfort, J.~U. Lee, D.~B. Newell, C.~A. Richter, Edge-state transport in graphene p-n junctions in the quantum hall regime, Physical Review B 92~(24) (2015) 241301.

\bibitem{tworzydlo2007valley}
J.~Tworzyd{\l}o, I.~Snyman, A.~Akhmerov, C.~Beenakker, Valley-isospin dependence of the quantum hall effect in a graphene p-n junction, Physical Review B—Condensed Matter and Materials Physics 76~(3) (2007) 035411.

\bibitem{trifunovic2019valley}
L.~Trifunovic, P.~W. Brouwer, Valley isospin of interface states in a graphene pn junction in the quantum hall regime, Physical Review B 99~(20) (2019) 205431.

\bibitem{myoung2020manipulation}
N.~Myoung, H.~Choi, H.~C. Park, Manipulation of valley isospins in strained graphene for valleytronics, Carbon 157 (2020) 578--582.

\bibitem{handschin2017giant}
C.~Handschin, P.~Makk, P.~Rickhaus, R.~Maurand, K.~Watanabe, T.~Taniguchi, K.~Richter, M.-H. Liu, C.~Schonenberger, Giant valley-isospin conductance oscillations in ballistic graphene, Nano letters 17~(9) (2017) 5389--5393.

\bibitem{myoung2024detecting}
N.~Myoung, T.~Song, H.~C. Park, Detecting strain effects due to nanobubbles in graphene mach--zehnder interferometers, physica status solidi (b) 261~(7) (2024) 2300379.

\bibitem{assouline2021excitonic}
A.~Assouline, M.~Jo, P.~Brasseur, K.~Watanabe, T.~Taniguchi, T.~Jolicoeur, D.~Glattli, N.~Kumada, P.~Roche, F.~Parmentier, et~al., Excitonic nature of magnons in a quantum hall ferromagnet, Nature Physics 17~(12) (2021) 1369--1374.

\bibitem{wei2018electrical}
D.~S. Wei, T.~Van Der~Sar, S.~H. Lee, K.~Watanabe, T.~Taniguchi, B.~I. Halperin, A.~Yacoby, Electrical generation and detection of spin waves in a quantum hall ferromagnet, Science 362~(6411) (2018) 229--233.

\bibitem{myoung2017conductance}
N.~Myoung, H.~C. Park, Conductance oscillations in chern insulator junctions: Valley-isospin dependence and aharonov-bohm effects, Physical Review B 96~(23) (2017) 235435.

\bibitem{groth2014kwant}
C.~W. Groth, M.~Wimmer, A.~R. Akhmerov, X.~Waintal, Kwant: a software package for quantum transport, New Journal of Physics 16~(6) (2014) 063065.

\end{thebibliography}

\end{document}